\documentstyle[psfig]{mn}

\title[PG1115+166 -- a long period DA+DB binary.]
{PG1115+166 -- a long period DA+DB binary.}
\author[P. F. L. Maxted et~al.]
       {P. F. L. Maxted$^{1,2}$,  M. R. Burleigh$^3$, T. R. Marsh$^2$ and
N. P. Bannister$^3$ \\
$^1$    School of Chemistry \& Physics, Keele University, Staffordshire,
ST5~5BG, UK \\
$^2$    University of Southampton, Department of Physics \& Astronomy,
        Highfield, Southampton, S017 1BJ, UK \\
$^3$    Department of Physics and Astronomy, University of Leicester,
University Road, Leicester LE1 7RH, UK
}
\date{Accepted ---
      Received --- }

\pagerange{\pageref{firstpage}--\pageref{lastpage}}
\pubyear{1999}

\newcommand{\Msolar}{\mbox{${\rm M}_{\odot}$}}
\newcommand{\Teff}{\mbox{${\rm T}_{\rm eff}$}}
\newcommand{\Rsolar}{\mbox{${\rm R}_{\odot}$}}
\newcommand{\kms}{\mbox{${\rm km\,s}^{-1}$}}
\begin{document}

\maketitle

\label{firstpage}

\begin{abstract}
 We present spectra of the DAB white dwarf PG\,1115+166. Radial velocity
measurements of the Balmer lines and the HeI\,6678\AA\ line show that this is
a binary white dwarf with a period of 30.09d in which the Balmer lines move
in anti-phase to the HeI line, i.e., PG\,1115+166 is a DA+DB binary. The
minimum masses of the stars are $M_{\rm DA} = 0.43 \pm 0.15\Msolar$ and
$M_{\rm DB} = 0.52 \pm 0.12\Msolar$. The separation of the stars is about
45\Rsolar, which is much smaller than a typical AGB progenitor of a white
dwarf, implying that there has been at least one common envelope phase in this
binary. Indeed, it is possible that this binary may have suffered up to three
mass transfer episodes -- two associated with the red giant phase prior to the
formation of each white dwarf and a third associated with the ``born-again''
red giant phase of the DB white dwarf.  PG\,1115+166 has the longest orbital
period of any post common envelope  white dwarf -- white dwarf binary found to
date. Published models for the formation of white dwarf -- white dwarf
binaries do not predict any white dwarfs with the combination of long orbital
period and high mass found in PG\,1115+166. We conclude that PG\,1115+166 is a
key object for testing models of binary star evolution and it may also be a
key object for our understanding of the formation of DB white dwarfs. We
outline the observational tests which can be applied to scenarios for the
formation of PG\,1115+166 and apply them to the simplest case of a single
common envelope phase. This suggests that some part of the internal energy
stored in the envelope of the AGB star, e.g., as ionized hydrogen, may have
contributed to the ejection of the common envelope, but there are several
unanswered questions concerning this simple scenario.

\end{abstract} 

\section{Introduction}
 The chemical composition of the atmosphere of a white dwarf star reflects a
balance between gravitational settling, which tends to produce a pure hydrogen
atmosphere,  and processes which ``pollute'' the atmosphere such as radiative
levitation, convection and accretion (Fontaine \& Wesemael 1987). This balance
will change during the lifetime of a white dwarf as it cools and this is seen
as a change in the relative numbers of hydrogen-rich (DA) and helium rich
(non-DA) white dwarfs as a function of effective temperature.  In white dwarfs
with effective temperatures $\Teff\sim$12\,--\,60\,000K, gravitational settling
dominates so  about 80\,percent of these white dwarfs have pure or nearly-pure
hydrogen atmospheres. White dwarf stars with helium-rich atmospheres may
appear as DO stars with He\,II lines ($\Teff \ga$45\,000K) or DB stars with
He\,I lines (30\,000K $\ga \Teff \ga $12\,000K).   The absence of non-DA white
dwarfs with effective temperatures between 30\,000K and 45\,000K is known as the
DB gap. The favoured explanation for the DB gap is that as the DO white dwarf
cools to 45\,000K, small amounts of hydrogen hidden in the atmosphere rise to
the surface, eventually masking the underlying helium. A hydrogen layer mass
of just $10^{-14}\Msolar$ is enough to make the white dwarf look like a DA.
Later, at $\sim 28\,000$K, the convection zone in the underlying helium layer
reaches the photosphere, mixing the hydrogen and helium so the white dwarf
re-appears as a helium-rich, DB white dwarf.  Those white dwarfs with thicker
hydrogen layers remain as DA stars in this temperature range. In this
scenario, $\sim20$\,percent of DA white dwarfs within the DB gap have thin
hydrogen layers which are sufficient to mask the underlying helium. 

The reason why $\sim 20$\,percent of white dwarfs are born with very thin
hydrogen envelopes is thought to be related to the born-again red giant
phenomenon in which some stars form hot white dwarfs with thick hydrogen
envelopes which undergo a late thermal pulse. This is caused by re-ignition of
the helium shell which results in the stars rapidly expanding to red giant
dimensions, e.g., Sakurai's Object (Duerbeck 2000). The remaining hydrogen is
quickly mixed into the helium layers and is burnt (Iben et~al.  1983).

 Those few white dwarfs which show a mixture of hydrogen and helium in their
atmosphere are particularly interesting in the light of this debate. About
20 percent of DB stars show traces of hydrogen in their atmospheres and are
classified as DBA (Shipman, Liebert \& Green 1987). Much rarer are stars which
show both strong hydrogen lines and strong helium lines. These  are classified
as DAB stars and only six of them have been identified. The prototype, GD\,323,
appears at the cool edge of the DB gap (\Teff = 28\,750K) and appears to have a
very thin hydrogen layer (10$^{-17}$\Msolar) although no model can yet explain
all the features of this peculiar object satisfactorily (Koester, Liebert \&
Saffer 1994).  MCT\,0128$-$3846 and MCT\,0453$-$2933 appear to be binaries
with a DA and a DB or DBA component (Wesemael et~al., 1994). Holberg, Kidder
\& Wesemael (1990) detected a weak HeI\,4471\AA\ feature in G104$-$27
(WD\,0612+177) but this was not confirmed by subsequent observations (Kidder
et~al. 1992).  HS\,0209+0832 is a DAB star in the middle of the DB gap which
shows strong HeI lines whose strength varies from year-to-year (Heber et~al.
1997) which may be explained by accretion onto the white dwarf from an
inhomogeneous interstellar cloud (Wolff et~al. 2000).  Finally, PG\,1115+166
was first reported as a DAB star by Burleigh et~al.  (2001). They note that
the H$\alpha$ line shows a variable radial velocity but were unable to
determine the orbital period. Bergeron \& Liebert (2002) have analysed the
optical spectrum of PG\,1115+166 and found an excellent fit to the spectrum by
assuming it is an unresolved DA+DB double degenerate binary.  Their fit to the
observed spectrum used  a  DA star with T$_{\rm eff}=22090$K, $\log g= 8.12$
combined with a helium-line DB star with T$_{\rm eff}=16210$K, $\log g= 8.19$.
They estimate that both stars have masses of about 0.7\Msolar.

 Those DAB stars which are binaries may be particularly interesting as they
may be examples of stars which have undergone two mass transfer episodes,
e.g., a common envelope phase (Iben \& Livio 1993). The number of such close
white dwarf pairs has increased rapidly in recent years. The majority of these
binaries were identified as low mass white dwarfs ($\la 0.45\Msolar$) from
analysis of their Balmer lines. White dwarfs of such low mass are not formed
in the standard picture of single star evolution within the lifetime of our
Galaxy. It now appears that the majority of these binaries are the result of a
common envelope phase in a star on the first giant branch (Marsh, Dhillon \&
Duck 1995). The properties of the binaries identified to-date, e.g., their
orbital periods, masses and mass ratios, are strong tests of models for the
formation of close binary white dwarfs and models for interacting binary stars
in general (Nelemans et~al. 2000, Nelemans et~al. 2001). Identifying DAB stars
 may be a promising way to find more massive close binary white dwarf stars
and so extend the range of masses over which observations can be used to test
models of interacting binary stars.

 In this paper we report radial velocity measurements of the Balmer lines and
HeI lines of PG\,1115+166. The Balmer lines show that this star is a binary
with a period of 30.09d and the helium lines move in anti-phase to the
hydrogen lines, so this is star is a DA+DB binary white dwarf. We also
consider the formation of PG\,1115+166.

\begin{table*}
\caption{\label{ObsTable} Summary of the spectrograph/telescope combinations
used to obtain spectra of PG\,1115+166 for this study. The slit width used in
each case is
approximately 1arcsec. }
\begin{tabular}{@{}lllcccl} 
&&&\multicolumn{1}{l}{No.of}&Sampling&Resolution& Spectral lines\\
Telescope & Date & Spectrograph& spectra&(\AA)&(\AA) &observed \\
\hline
\noalign{\smallskip}
INT & 1996 Feb & IDS 235mm & 2& 1.59 & 3.2 & H$\beta$ -- H$_9$  \\
WHT & 1997 Nov & ISIS Blue  & 2& 0.22 & 0.7 & H$\beta$, H$\gamma$  \\
WHT & 1997 Nov & ISIS Red   & 2& 0.40 & 0.8 & H$\alpha$, He\,I6678  \\
INT & 1999 Feb & IDS 500mm & 3& 0.39 & 0.9 & H$\alpha$, HeI6678  \\
WHT & 1999 Apr & ISIS Blue  & 1& 0.22 & 0.9 & H$\beta$ -- H$\gamma$        \\
WHT & 1999 Apr & ISIS Red   & 2& 0.79 & 1.6 & HeI\,6678, HeI\,7065  \\
WHT & 1999 Dec & ISIS Blue  & 1& 0.44 & 1.0 & H$\beta$ -- H$_9$  \\
INT & 2000 Apr & IDS 500mm & 6& 0.39 & 0.9 & H$\alpha$, HeI\,6678  \\
WHT & 2001 Jan & ISIS Blue  & 6& 0.21 & 0.4 & H$\beta$  \\
WHT & 2001 Jan & ISIS Red   & 6& 0.40 & 0.8 & H$\alpha$, HeI\,6678  \\
INT & 2001 Feb & IDS 500mm & 1& 0.39 & 0.9 & H$\alpha$, HeI\,6678   \\
WHT & 2001 Mar & ISIS Blue  & 3& 0.45 & 1.2 & H$\gamma$ -- H$\epsilon$  \\
WHT & 2001 Mar & ISIS Red   & 3& 0.40 & 0.8 & H$\alpha$, HeI\,6678  \\   
WHT & 2001 Apr & ISIS Red   & 1& 0.40 & 0.8 & H$\alpha$, HeI\,6678   \\
WHT & 2002 Jan & ISIS Blue  & 4& 0.22 & 0.7 & H$\gamma$, H$\delta$  \\
WHT & 2002 Jan & ISIS Red   & 4& 0.40 & 0.8 & H$\alpha$, HeI\,6678  \\
\hline
\noalign{\smallskip}
\end{tabular}
\end{table*}

\begin{table}
\caption{\label{RVTable}Radial velocities measured from the Balmer lines 
and the HeI lines in PG\,1115+166.} 
\begin{tabular}{@{}lrlrl}
HJD& \multicolumn{1}{l}{Radial Velocity} & \multicolumn{1}{l}{Line} &
 \multicolumn{1}{l}{Radial Velocity} \\
 -2450000  & (km\,s$^{-1}$) & & (km\,s$^{-1}$) & Line \\
\noalign{\smallskip}
~114.6490 &  118$\pm$107  & H$\delta$ &   52$\pm$ 69& 4026 \\
~114.6490 &   90$\pm$ 38  & H$\beta$  &   40$\pm$ 72& 4471 \\
~114.6490 &  $-$86$\pm$ 50  & H$\gamma$ &             &      \\
~114.6621 &   79$\pm$ 46  & H$\gamma$ &$-$71$\pm$ 61& 4026 \\
~114.6621 &   44$\pm$ 96  & H$\delta$ &$-42 \pm$ 65 & 4471 \\
~114.6621 &   33$\pm$ 32  & H$\beta$  &             &      \\
~778.7425 &   48$\pm$  7  & H$\alpha$ &   62$\pm$ 22& 6678 \\
~778.7425 &               &           &             &      \\
~778.7474 &   44$\pm$ 10  & H$\beta$  &   18$\pm$ 30& 4471 \\
~778.7474 &   52$\pm$ 18  & H$\gamma$ &             &      \\
~778.7521 &   57$\pm$  7  & H$\alpha$ &    6$\pm$ 21& 6678 \\
~778.7521 &               &           &             &      \\
~778.7570 &   59$\pm$ 10  & H$\beta$  &   6$\pm$ 28 & 4471 \\
~778.7570 &   73$\pm$ 16  & H$\gamma$ &             &      \\
1242.4790 &    1$\pm$  8  & H$\alpha$ &             &      \\
1242.4790 &               &           &   33$\pm$ 26& 6678 \\
1242.6647 &$-$16$\pm$ 10  & H$\alpha$ &             &      \\
1242.6647 &               &           &   61$\pm$ 31& 6678 \\
1243.5684 &               &           &   81$\pm$ 30& 6678 \\
1243.5684 &   12$\pm$ 10  & H$\alpha$ &             &      \\
1264.4391 &               &           &   45$\pm$ 11& 7065 \\
1264.4391 &               &           &   80$\pm$ 16& 6678 \\
1265.4375 &   19$\pm$ 12  & H$\gamma$ &   37$\pm$ 20& 4471 \\
1265.4375 &   29$\pm$  7  & H$\beta$  &   59$\pm$  9& 7065 \\
1265.4375 &               &           &   61$\pm$ 13& 6678 \\
1534.7763 &   42$\pm$ 13  & H$\delta$ &   20$\pm$ 12& 4026 \\
1534.7763 &   25$\pm$  6  & H$\beta$  &   42$\pm$ 18& 4471 \\
1534.7763 &   24$\pm$ 11  & H$\gamma$ &             &      \\
1646.5221 &   94$\pm$  7  & H$\alpha$ &    3$\pm$ 23& 6678 \\
1647.3903 &               &           &   23$\pm$ 43& 6678 \\
1647.5046 &   52$\pm$ 14  & H$\alpha$ &    5$\pm$ 44& 6678 \\
1654.5432 &   50$\pm$  8  & H$\alpha$ &   66$\pm$ 23& 6678 \\
1656.4903 &   23$\pm$  8  & H$\alpha$ &   99$\pm$ 24& 6678 \\
1657.5243 &   15$\pm$  8  & H$\alpha$ &   16$\pm$ 25& 6678 \\
1924.7813 &   34$\pm$  6  & H$\alpha$ &   48$\pm$ 20& 6678 \\
1924.7814 &   15$\pm$ 12  & H$\beta$  &             &      \\
1924.7884 &   34$\pm$  6  & H$\alpha$ &   36$\pm$ 20& 6678 \\
1924.7885 &   39$\pm$ 12  & H$\beta$  &             &      \\
1924.7954 &   23$\pm$  7  & H$\alpha$ &   27$\pm$ 22& 6678 \\
1924.7956 &   42$\pm$ 13  & H$\beta$  &             &      \\
1924.8024 &   30$\pm$  7  & H$\alpha$ &   22$\pm$ 25& 6678 \\
1924.8027 &   29$\pm$ 13  & H$\beta$  &             &      \\
1925.7458 &   29$\pm$  4  & H$\alpha$ &   43$\pm$ 14& 6678 \\
1925.7459 &  40$\pm$  8  & H$\beta$  &             &      \\
1925.7598 &  33$\pm$  5  & H$\alpha$ &   29$\pm$ 15& 6678 \\
1925.7599 &   8$\pm$  9  & H$\beta$  &             &      \\
1946.6078 &  61$\pm$ 17  & H$\alpha$ &             &      \\
1976.6739 &  71$\pm$ 14  & H$\alpha$ &             &      \\
1976.6740 &              &           &   32$\pm$ 47& 4471 \\
1976.6740 & $ -6\pm$ 38  & H$\delta$ &             &      \\
1976.6740 &              &           & $-$3$\pm$ 35& 4026 \\
1976.6740 &  56$\pm$ 28  & H$\gamma$ &             &      \\
1976.6879 &  74$\pm$ 10  & H$\alpha$ &             &      \\
1976.6880 &              &           &   18$\pm$ 27& 4026 \\
1976.6880 &              &           &$-$32$\pm$ 38& 4471 \\
\multicolumn{3}{l}{\it Continued \dots}
\end{tabular}
\end{table}
\begin{table}
\contcaption{}
\begin{tabular}{@{}lrlrl}
HJD& \multicolumn{1}{l}{Radial velocity} & \multicolumn{1}{l}{Line} &
\multicolumn{1}{l}{Radial Velocity} \\
-2450000  & (km\,s$^{-1}$) & & (km\,s$^{-1}$) & Line \\
\noalign{\smallskip}
1976.6880 &  81$\pm$ 31  & H$\delta$ &             &      \\
1976.6880 &  76$\pm$ 21  & H$\gamma$ &             &      \\
1976.7090 &  25$\pm$ 20  & H$\gamma$ &  $-18\pm$ 36& 4471 \\
1976.7090 & 113$\pm$ 35  & H$\delta$ & $-$5$\pm$ 26& 4026 \\
1976.7090 &  71$\pm$  8  & H$\alpha$ &             &      \\
2278.7485 &  77$\pm$  6  & H$\alpha$ & 11 $\pm$ 19 & 6678 \\
2278.7627 &  76$\pm$  6  & H$\alpha$ &$-4 \pm$ 22  & 6678 \\
2278.7797 &  73$\pm$  6  & H$\alpha$ &$28 \pm$ 22  & 6678 \\
2278.7938 &  71$\pm$  6  & H$\alpha$ &$-9 \pm$ 22  & 6678 \\
2278.7485 &  61$\pm$ 11  & H$\gamma$ &  $19\pm$19  & 4471 \\
2278.7627 &  51$\pm$ 17  & H$\gamma$ &  $29\pm$28  & 4471 \\
2278.7797 &  72$\pm$ 18  & H$\gamma$ & $-12\pm$31  & 4471 \\
2278.7939 &  67$\pm$ 19  & H$\delta$ &             &      \\
2278.7627 &  57$\pm$ 32  & H$\delta$ &             &      \\
2278.7939 &  36$\pm$ 37  & H$\delta$ &             &      \\
\end{tabular}
\end{table}

\section{Observations and Reductions}

 PG\,1115+166 was first identified as a DAB star in February 1996 as part of a
survey to find low mass white dwarfs (Moran 1999). It was re-observed several
times thereafter as part of the programme at Southampton to find and
characterize close binary white dwarfs. The Leicester group observed
PG\,1115+166 independently in April 1999 as a part of a programme to find
white dwarfs with helium-rich atmospheres in the DB gap by observing hot white
dwarfs undetected by EUV surveys and also identified it as a DAB star
(Burleigh et~al. 2001)

Observations of PG\,1115+166 have been obtained with the intermediate
dispersion spectrograph (IDS) on the 2.5m Isaac Newton Telescope (INT) and the
dual-beam ISIS spectrograph on the 4.2m William Herschel Telescope (WHT) both
on the Island of La Palma.  A variety of gratings have been used to cover a
variety of wavelengths at different times, as detailed in
Table~\ref{ObsTable}.  The exposure times used were typically 10--30\,minutes.
Spectra of an arc lamp are taken before and after each target spectrum with
the telescope tracking the star. None of the CCDs used showed any structure in
unexposed images, so a constant bias level determined from a clipped-mean
value in the over-scan region was subtracted from all the images. Sensitivity
variations were removed using observations of a tungsten calibration lamp.

\begin{figure*}
\caption{The spectrum of PG\,1115+166.
\label{SpecFig}}
\psfig{file=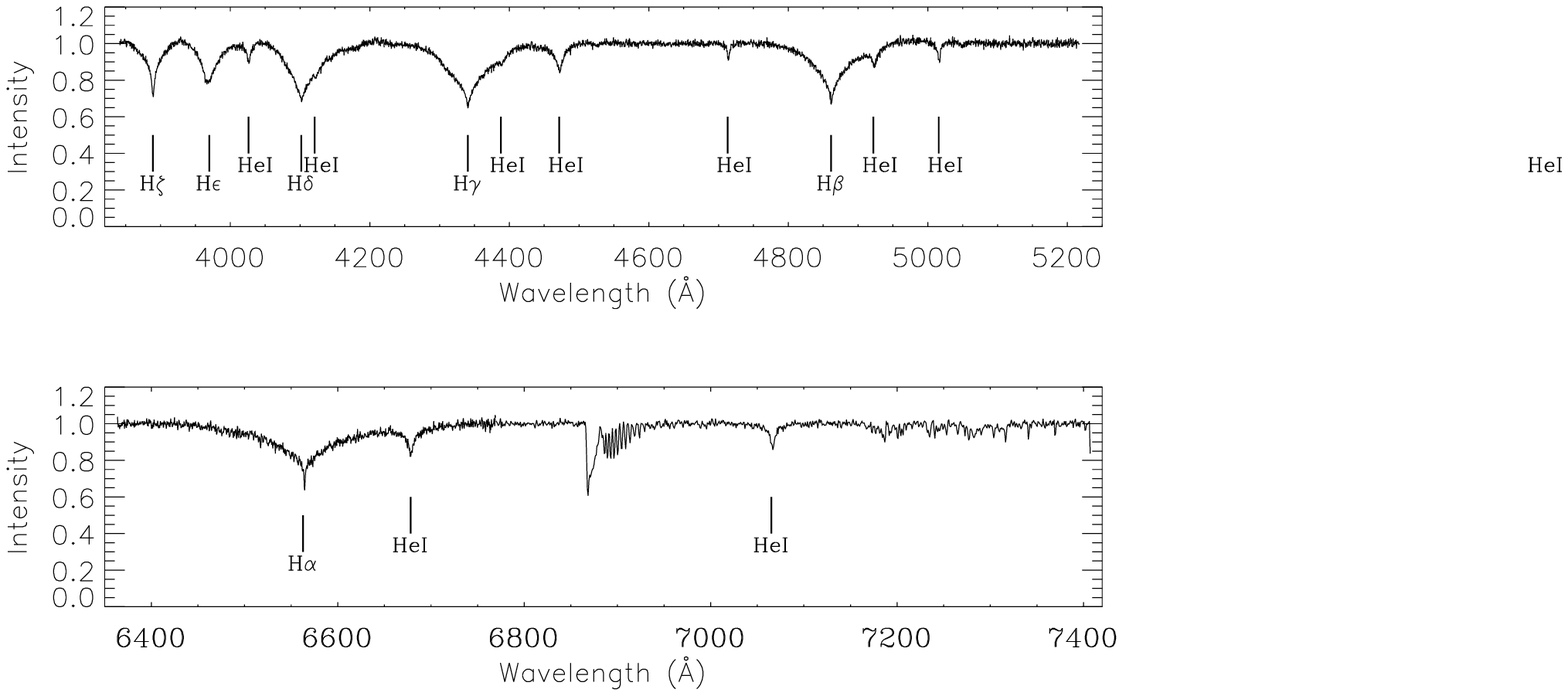,width=1.0\textwidth}
\end{figure*}

 Extraction of the spectra from the images was performed automatically using
optimal extraction to maximize the signal-to-noise of the resulting spectra
(Marsh 1989).  The arcs associated with each stellar spectrum were extracted
using the profile determined for the stellar image to avoid possible
systematic errors due to tilted arc lines. The wavelength scale was determined
from a polynomial fit to measured arc line positions and the wavelength of the
target spectra interpolated from the calibration established from the
bracketing arc spectra. Uncertainties on every data point calculated from
photon statistics are rigorously propagated through every stage of the data
reduction. The normalized spectrum of PG\,1115+166 is shown in
Fig.~\ref{SpecFig} in two sections. The spectra used to produce this figure
were the 1999 Dec WHT spectrum, the 1999 Apr WHT spectrum and the sum of the
three WHT spectra from  2002 Jan. No attempt has been made to remove the
telluric features near 6900\AA\ and 7300\AA.

\section{Analysis}
\subsection{Radial velocities.}

  To measure the radial velocities we used least-squares fitting of a model
line profile. This model line profile is the summation of one or more Gaussian
profiles with different widths and depths but with a common central position
which varies between spectra and is convolved with a Gaussian profile of an
appropriate  width to model the resolution of each spectrum.  

One must be careful to allow for the asymmetry of the line due to pressure
shifts, particularly when dealing with the wings of the higher order Balmer
lines (Grabowski, Halenka \& Madej 1987) or the HeI lines (Beauchamp, Wesemael
\& Bergeron 1997). To account for these asymmetries in the Balmer lines, we add
a 3rd or  4th order polynomial in the model line profile.  We also exclude
data from the fitting process which is more than 1000\,km\,s$^{-1}$ from the
rest wavelength of the lines measured. This reduces both the effect of the
pressure shift and the possibility of blending with other lines. The radial
velocity measurement is determined primarily by the centre of the line and the
fitting process is identical for all lines, so any remaining pressure shifts
would only be seen as an offset of a few km\,s$^{-1}$ between the radial
velocities measured for different Balmer lines. This is small compared to the
typical uncertainties in the measured radial velocities.

 The pressure shifts in the much weaker HeI lines are harder to deal with, so
we simply report the radial velocities measured using a single Gaussian profile
and a low-order polynomial for these lines and include a warning here that
there are systematic shifts between the velocities measured for different HeI
lines. 

 We first normalize the spectra using a polynomial fit to the continuum either
side of the line of interest. We use a least-squares fit to one of the spectra
to determine the model line profile. A least squares fit of this profile to
each spectrum in which the position of the line and the polynomial
coefficients are the only free parameters gives the measured radial velocities
in Table~\ref{RVTable}. We tried three different initial values for the radial
velocity in the least-squares minimization to ensure we found the optimum
value of the radial velocity and excluded the results for any spectra where
the optimum value could not be clearly identified.

 The uncertainties on each radial velocity measurement are calculated by
propagating the uncertainties on every data point in the spectra right through
the data reduction and analysis. There are inevitably systematic errors in our
radial velocity measurements, particularly since we are using spectra obtained
with a variety of instruments. Maxted et~al. (2000) found that these
systematic differences are no more than $\sim 1$\,km\,s$^{-1}$ for the
majority of the spectrographs used in this study. To allow for this, we
include an additional uncertainty per radial velocity measurement equivalent
to 1/40 of the resolution, typically about 1km\,s$^{-1}$. This has been added
in quadrature to the uncertainties given in Table~\ref{RVTable}. The radial
velocity measurements of the Balmer lines are clearly variable. Of the HeI
line measurements, only the measurements of the HeI\,6678 lines are
sufficiently numerous and accurate to show any sign of variability. There are
too few accurate radial velocity measurements of the other HeI lines for this
variability to be apparent. To quantify this we calculated their weighted mean
radial velocity, which is the best estimate of the radial velocity assuming
this quantity is constant. We then calculated the $\chi^2$ statistic for this
``model'', i.e., the goodness-of-fit of a constant to the observed radial
velocities. We can then compare the observed value of $\chi^2$ with the
distribution of $\chi^2$ for the appropriate number of degrees of freedom and
find the probability of obtaining the observed value of $\chi^2$ or higher
from random fluctuations of constant value, $p=0.023$. This suggests that the
HeI\,6678 lines are likely to be variable, but the evidence from this test is
not conclusive. 

\subsection{The orbital period}
 We used the 57 radial velocities measured from the Balmer lines to establish
the orbital period. We used least-squares to fit sine waves at 500\,000 trial
frequencies uniformly distributed over the range 0 to 10 cycles per day and
recorded the value of chi-squared for each trial frequency. Low values of
chi-squared occur for orbital periods $P \approx 31$d with the lowest value
of chi-squared being  68.3 for $P=30.0873$d. The next most likely orbital
period is near the one cycle-per-day alias of this period, $P=1.0316$d, which
has a chi-squared value of 87.3. 

 In order to test the robustness of our period determination, we randomly
selected 50 of the radial velocity measurements and fitted sine waves at
100\,000 trial frequencies uniformly distributed over the range 0 to 1.2
cycles per day and recorded the value of the period at which the minimum
value of chi-squared occurs. We repeated this process 1000 times and found
that in 973 trials this reduced dataset produced a minimum value of
chi-squared at an orbital period near 30.09d. Most of the other trials favoured
orbital periods near 1 day which, as we have seen, gives a poor fit to the
complete data set. 

 As an alternative test, we used a Monte Carlo simulation to estimate how
often  a binary with an orbital period of 1.0316d will appear to have an orbital
period near 30.09d given radial velocity meausurements with the same temporal 
sampling and accuracy as those presented here. We used a least-sqaures fit to
our data to find the optimum period near 1.0316d and the corresponding
semi-amplitude of a sine wave fit by least squares. We generated 1000 synthetic
data sets with the same period and semi-amplitude as this fit and with the
same temporal sampling as our data. We added noise to the synthetic data using
a psuedo-random normal deviates multiplied by the uncertainties associated
with each datum. For each of these synthetic data sets we fitted sine waves at
100\,000 trial frequencies uniformly distributed over the range 0.01 to 1.2
cycles per day and recorded the value of the period at which the minimum value
of chi-squared occurs. We also recorded  the best values of chi-squared which
occur within 10\,percent of $P=1.0316$d and $P=30.09$d. Of the 1000 trials,
983 trials resulted in a minimum value of chi-squared near 1.0316d and only 17
gave a minimum value of chi-squared at a period near 30.09d. Furthermore,  in
all the trials for which $P\approx 30.09$d gave a lower value of chi-squared,
the difference in the minimum chi-squared value for  $P\approx 1.0316$d and
$P\approx 30.09$d was usually only 1--3 and was always less than 8. In
summary, this test suggests that there is a less than 1/1000 chance of a
binary with $P=1.0316$d being incorrectly identified as a binary with
$P=30.09$d from our data provided the least-squares fits of sine waves at the
two periods result in a difference in chi-squared of 8 or more. The difference
in chi-squared for our actual data is 19, so this condition is easily
satisfied in this case.

These simple test both suggests that period determination is robust and that
we can be confident that the orbital period is $P\approx 30.09$d.

\subsection{Spectroscopic orbit}
\begin{figure}
\caption{Radial velocities measured from the Balmer lines (filled symbols) and
the HeI\,6678\AA\ line (open symbols) of PG\,1115+166 as a function of orbital
phase. The sine waves fit by least-squares described in the text are also
shown.\label{RVFig}}
\psfig{file=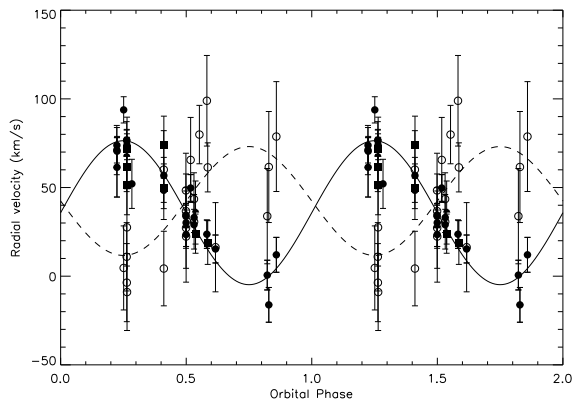,width=0.45\textwidth}
\end{figure}

 If PG\,1115+166 is genuinely a binary  DA+DB white dwarf we expect that the
HeI lines should move in anti-phase to the Balmer lines. We investigated this
scenario using a least-squares fit of a sine wave of the form $v_i =
\gamma_{\rm H} + K_{\rm H}\sin(2\pi(T_i-T_0)/P$  to the  measured radial
velocities, $v_i$, for all the Balmer lines and a sine wave of the form $v_i =
\gamma_{\rm HeI} + K_{\rm HeI}\sin(2\pi(T_i-T_0)/P$ to the 23 measured radial
velocities for the HeI\,6678 line. We used both data sets in a simultaneous
fit to all these data, an approach which has the merit of using all the
available data to determine the parameters $T_0$ and $P$. The results are
given in Table~\ref{RVFitTable} and are shown in Fig.~\ref{RVFig}. We have
excluded data for the Balmer lines with uncertainties larger than 20\,km/s and
data for the HeI\,6678 line with uncertainties larger than 40\,km/s from
Fig.~\ref{RVFig} for clarity. The value of $K_{\rm HeI}$ derived is about 4
times greater than its uncertainty, so we can be fairly sure that the
HeI\,6678 line moves in anti-phase to the Balmer lines as expected for a DA+DB
binary. We can certainly rule out the possibility that PG\,1115+166 is a
single DAB star since that would require the  HeI lines and the Balmer lines
to move in phase, which is clearly not the case. Also given  in
Table~\ref{RVFitTable} are the results of a similar least-squares fit
including the data for all the HeI lines. This reduces the uncertainty in the
value of $K_{\rm HeI}$, but we prefer the results of fitting only the
HeI\,6678 line until the systematic errors introduced by pressure shifts in
HeI lines is better understood. In any case, the results of the two
least-squares fits are nearly identical.

\begin{table}
\caption{Results of least-squares sine wave fits to the measured radial
velocities for PG\,1115+166.\label{RVFitTable}}
\begin{tabular}{@{}lrr}
\noalign{\smallskip}
   & \multicolumn{1}{l}{HeI\,6678 only} & \multicolumn{1}{l}{All HeI lines}\\
$\gamma_{\rm H}$ (\kms)  &      35.8 $\pm$  2.3  &  35.7 $\pm$  2.3 \\
K$_{\rm H}$ (\kms)       &      36.6 $\pm$  2.4  &  36.6 $\pm$  2.4 \\
$\gamma_{\rm HeI}$ (\kms)&      40.6 $\pm$  4.7  &  37.2 $\pm$  3.6 \\
K$_{\rm HeI}$ (\kms)     &     $-30.7 \pm$  7.4  & $-$30.7 $\pm$ 5.4 \\
T$_0$(HJD)               &2451909.77 $\pm$  0.41  & 2451909.76 $\pm$  0.40 \\
P        (days)          &    30.089 $\pm$  0.016 & 30.088 $\pm$  0.016 \\
N                        &   80 & 100    \\
$\chi^2$                 & 88.14   & 100.53\\
\end{tabular}
\end{table}

 The mass ratio implied by the parameters in Table~\ref{RVFitTable} is
$q=M_{\rm DA}/M_{\rm DB} = 0.84\pm 0.21$. This is consistent with the
conclusion of Bergeron \& Liebert (2002) that the masses of the two stars are
similar. We also find $M_{\rm DA}\sin^3 i = 0.43\pm0.15\Msolar$, $M_{\rm
DB}\sin^3 i = 0.52\pm0.12\Msolar$ and $a\sin i = 40 \pm 5\Rsolar$, where $i$
is inclination of the binary and $a$ is the separation of the stars. The
masses measured by Bergeron \& Liebert suggest that the inclination of the
binary is approximately 60$^{\circ}$ and that the stars are separated by about
45\Rsolar.

\section{The formation of PG\,1115+166}
 The basic mechanism for the formation of most white dwarfs is fairly well
understood. They are, in general, remnants of intermediate mass stars
(1--8\Msolar) which have evolved through the asymptotic giant branch (AGB)
phase. Extreme mass loss towards the end of this phase exposes the 
core of the AGB star, which cools to form a white dwarf. Stars with larger
initial masses are expected to form more massive white dwarfs and there is
some observational evidence for this initial-final mass relation (IFMR;
Weidemann 2000).  White dwarfs with masses of 0.7\Msolar\  are expected to
form from stars with initial masses of about 3\Msolar. 

 Models of stars on the AGB are uncertain, but the radius of a star towards
the end of this phase is several hundred solar radii (Girardi et~al. 2000).
This is much larger than the current separation of the stars so it appears
that the orbit of PG\,1115+166 is now substantially smaller than it was prior
to the formation of the white dwarfs. Orbital shrinkage in long period
binaries is explained by the ``common envelope'' scenario. In this scenario,
the expanding red giant star comes into contact with its Roche lobe and begins
to transfer mass to its companion star. This mass transfer is highly unstable,
so a common envelope forms around the companion and the core of the red
giant. The drag on the companion orbiting inside the common envelope  leads to
extensive mass loss and dramatic shrinkage of the orbit (Iben \& Livio 1993).
We  expect the orbit of a post-CE binary to be circular, and have assumed this
to be so in the case of  PG\,1115+166 throughout this paper. Much higher
quality data than that presented here would be required to verify this
assumption.

\begin{figure}
\caption{PG\,1115+166 in the mass-period plane compared to the models of
Iben, Tutukov \& Yungelson (1997). The figure is taken from Saffer, Livio \&
Yungelson (1998) and shows the predicted number of observable white
dwarf\,--\,white dwarf binaries as a function of orbital period and mass as a
gray scale images. We have assumed an uncertainty of 0.05\Msolar\
in the mass for this figure. \label{SLYFig}}
\psfig{file=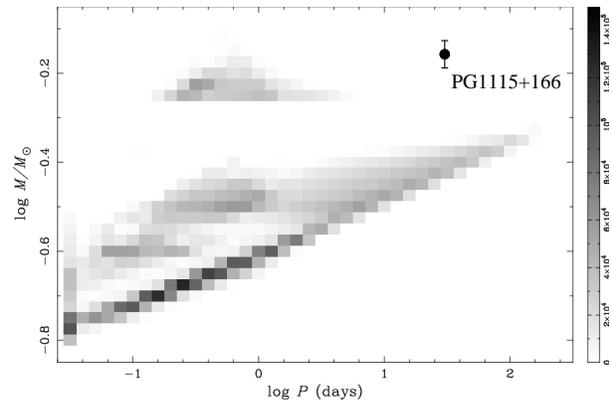,width=0.45\textwidth} \end{figure}

 Common envelope evolution is poorly understood so it is useful to consider
the limits imposed on the common envelope (CE) phase by the observed
properties of PG\,1115+166. Firstly, we note that published models for the
population of white dwarf binaries do not predict any systems with the
combination of high mass and long period seen in PG\,1115+166. This is shown
in Fig.~\ref{SLYFig} for the models of Iben, Tutukov and Yungelson (1997), but
the same applies to the models of Nelemans (2001) and Han (1998). It is
primarily the combination of high mass and long orbital period which is not
explained by these models. A full exploration of all the possible formation
scenarios for PG\,1115+166 is beyond the scope of this paper, not least
because there may have been up to three mass transfer episodes -- two
associated with the red giant phase prior to the formation of each white dwarf
and a third associated with the born-again red giant phase of the DB white
dwarf.  These mass transfer episodes may not have been CE phases but may
instead have involved stable mass transfer, i.e.,  an Algol-like phase.  In
this section we simply outline the observational constraints which can be
applied to these scenarios and apply these constraints to the simplest case of
a single CE phase.

\begin{figure}
\caption{The relationship between the initial masses of the progenitors of the
DB star ($M^{\prime}_1$) and the DA star ($M^{\prime}_2$) implied by the
difference in cooling ages of $(1.6\pm0.1)\times10^8$y based on the models of
Girardi et~al., (2000) for Z=0.019 (solid lines) and Z=0.004 (dashed lines).
\label{M1M2Fig}}
\psfig{file=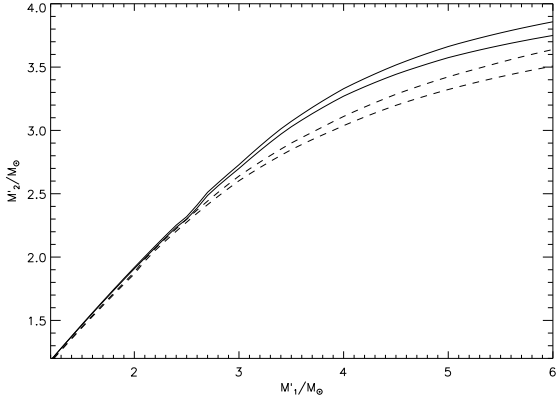,width=0.45\textwidth}
\end{figure}

Apart from the present day masses of the white dwarfs ($\sim 0.7\Msolar$) and
their current separation ($a_f \approx 45\Rsolar$), we also know the cooling
ages of the stars are $6\times10^7$y for the DA white dwarf and
$2.2\times10^8$y  for the DB white dwarf (Bergeron \& Liebert 2002). The
cooling ages are the time elapsed since the AGB phase. If the progenitors of
the white dwarfs were formed at the same time, this imposes a limit on the
initial masses of the progenitor stars as follows. If we assume an initial
mass for the progenitor of the DB star, $M^{\prime}_1$, we can use an
appropriate stellar model to find the lifetime of this star, $\tau_1$, i.e.,
the time taken  since it formed for the star to reach the end of AGB phase. We
then know that the progenitor of the DA star will take a further
$2.2\times10^8\rm y - 6\times10^7\rm y = 1.6\times10^8$y to reach the AGB
phase. We can then use grid of stellar models to find the initial mass,
$M^{\prime}_2$, of a star whose lifetime is $\tau_2 = \tau_1 +
1.6\times10^8$y, i.e, the mass of the progenitor of the DA star. The
relationship  between these initial masses is shown in Fig.~\ref{M1M2Fig}
based on the models of Girardi et~al. (2000) for Z=0.019 (solid lines) and
Z=0.004 (dashed lines). Although the stars have not evolved as single stars,
this argument regarding the cooling age difference between the stars remains
valid because the AGB phase is much shorter than the lifetime of the star
which precedes it, i.e., the stars spend most of their lives evolving as
single stars. However, we should be aware that the IFMR may not apply to the
white dwarfs in PG\,1115+166 in the sense that a star may produce a white
dwarf which is less massive than expected if it is involved in a CE phase
which interrupts the growth of the future white dwarf forming in its core.
Also, this constraint may not be valid if there has been significant
accretion onto one of the stars. Similarly,  if the born-again red giant phase
of the DB star was short, the DB white dwarf may not have been re-heated to
the temperatures expected for a white dwarf emerging from its first AGB phase.
In this case  the cooling age does not measure either the time since the AGB
phase, nor the time since the born-again read giant phase, but is somewhere
between these two timescales.

 The observation that one of the stars is a DB white dwarf is
also a constraint on possible  formation scenarios for  PG\,1115+166. Any
feasible scenario must allow for the DB white dwarf to undergo a born-again
red giant phase or include an alternative explanation for how this star came
to be hydrogen deficient. A stronger constraint, perhaps, is how this star
avoids accreting hydrogen onto it surface despite appearing to be older than
its companion which has, presumably, lost large quantities of hydrogen rich
material. 

 With this information we consider the simplest case of a single CE phase in
which the AGB star progenitor of the DA star with mass $M_2$ and radius $R_2$
interacts with the DB star when it is close to or exceeds its Roche lobe. Note
that $M_2$ will be less than $M^{\prime}_2$ if any mass loss occurs prior to
the interaction. However, this mass loss cannot have been very extensive in
this scenario because the DB white dwarf cannot have gained more than $\la
10^{-14}\Msolar$ so we assume $M_2 = M_2^{\prime}$. If the initial semi-major
axis  of the stars' orbit is $a_i$ and the current separation is $a_f$, the
change in the orbital binding energy is 
\[
\Delta E_{\rm orb} = \frac{GM_2M_{\rm DB}}{2a_i} - \frac{GM_{\rm DA} 
M_{\rm DB}}{2a_f} 
\]
 Some fraction of this  energy $\alpha_{\rm CE}$ is used to expel the envelope
of the AGB star whose binding energy is given by
\[
       E_{\rm env} = -\frac{GM_2 (M_2-M_{\rm DA})}{\lambda R_1}
\]
So from the definition of $\alpha_{\rm CE}$ we have
\[
 \alpha_{\rm CE} \lambda \left(\frac{GM_{\rm DA} M_{\rm DB}}{2a_f} -
\frac{GM_2M_{\rm DB}}{2a_i}\right) =  \frac{GM_2 (M_2-M_{\rm DA})}{R_1}
\]

 If the orbit prior to the interaction has an eccentricity $e$, then the $R_1
= r_L a_i (1-e)$ at the time of interaction, where $r_L$ is the radius of the
Roche lobe relative to the separation of the stars. This assumes that the
interaction occurs at periastron, which is reasonable given that the orbital
period ($\sim 10$y) is much shorter than the evolutionary timescale of an AGB
star. 

\begin{figure}
\caption{The efficiency of the common envelope phase, $\alpha_{\rm
CE}\lambda$, as a function of the
radius, $R_2$ and the mass $M_2$ of the AGB star progenitor of the DA white
dwarf in PG\,1115+166.
Solid lines correspond to values of $\alpha_{\rm CE}\lambda$ as marked.
The effect of a change of 0.02\Msolar\  in the current masses of the white
dwarfs is illustrated by the dotted lines either side of the solid lines
 \label{LambdaFig}}
\psfig{file=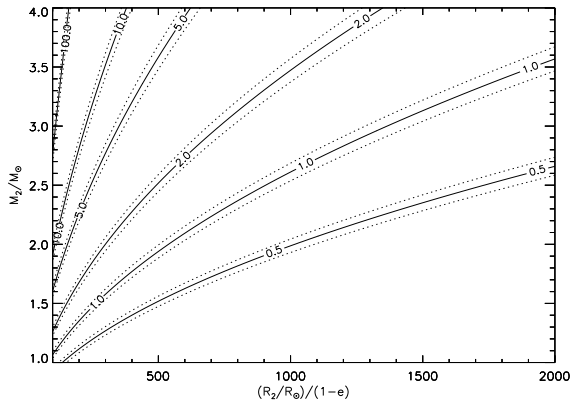,width=0.45\textwidth}
\end{figure}

 In Fig.~\ref{LambdaFig} we show lines of constant $ \alpha_{\rm CE} \lambda$
as a function of $M_2$ and $R_2/(1-e)$ for $M_{\rm DA}=M_{\rm DB}=0.7\Msolar$
and $a_f = 45\Rsolar$. The maximum radius of an AGB star with an initial mass
of $\sim 3 \Msolar$ is a few hundred solar radii (Bloecker 1995), so unless
the eccentricity of the orbit prior to the CE phase was quite extreme ($e \ga
0.5$) and/or the initial mass of the AGB star was lower than suggested by the
IFMR ($\approx 2\Msolar$), we see that $ \alpha_{\rm CE}\lambda \ga 1.0$. This
requires a large value of $\lambda$ because $\alpha_{\rm CE} < 1$ in the
absence of other sources of energy to eject the envelope.  This strong limit
on  $\lambda $ is a consequence of the combination of long orbital period and
high mass in PG\,1115+166.  Large values of $\lambda$ for AGB stars are
predicted by Dewi \& Tauris (2000) provided the calculation of the binding
energy of the envelope includes the internal energy stored in the envelope,
e.g., as ionized hydrogen. This implicitly assumes that this potential energy
is efficiently converted into the kinetic energy of the envelope during the
common envelope phase. Several issues arising from  this scenario remain to be
explored, e.g., how does the DB star avoid gaining more than $10^{-14}\Msolar$
of hydrogen during the CE phase, why was there no CE phase when the progenitor
of the DB star went through its AGB phase and  where does the born-again red
giant phase fit into this scenario?  It is beyond the scope of this paper to
tackle these questions, but they do demonstrate that our observations of
PG\,1115+166 provide a challenging test of models for binary star evolution
and the formation of DB white dwarfs.

\section{Conclusion}

  We have used radial velocity measurements of the Balmer lines and 
HeI lines of PG\,1115+166 to show that this star is a DA+DB binary white dwarf
with an orbital period of 30.09d.  Published models for the
formation of white dwarf --  white dwarf binaries do not predict any white
dwarfs with the characteristics of PG\,1115+166. We consider the observational
constraints on scenarios for the  formation of PG\,1115+166. We apply these to
the simplest case of a single common-envelope phase. We find that this
requires that some part of the  internal energy in the envelope of the
AGB star contributes to the ejection of the common envelope, but several
questions regarding this scenario remain to be answered. We conclude that
PG\,1115+166 is a key object for testing models of binary star evolution. It
may also  be a key object for our understanding of the formation of DB white
dwarfs.

\section*{Acknowledgments}
 PFLM would like to thank Jacco van~Loon for his helpful discussions regarding
AGB stars and Phillip Podsiadlowski for sharing his thoughts on the formation
of PG\,1115+166. The Isaac Newton Telescope is operated on the island of La
Palma by the Isaac Newton Group in the Spanish Observatorio del Roque de los
Muchachos of the Instituto de Astrofisica de Canarias.

\label{lastpage}

\end{document}